\documentclass[12pt]{article}

\usepackage{amsmath,amssymb,amsthm}

\newcommand{\real}{\mathbb{R}}

\newcommand{\vj}{\boldsymbol{j}}
\newcommand{\vv}{\boldsymbol{v}}

\newcommand{\vs}{\boldsymbol{s}}

\newcommand{\Gammav}{\varGamma}

\begin{document}

\title{\bf  
Property of Zero-energy Flows and 
\break 
Creations and Annihilations of Vortices 
\break
in Quantum Mechanics 
}

\author{Tsunehiro Kobayashi\footnote{E-mail: 
kobayash@a.tsukuba-tech.ac.jp} \\
{\footnotesize\it Department of General Education 
for the Hearing Impaired,}
{\footnotesize\it Tsukuba College of Technology}\\
{\footnotesize\it Ibaraki 305-0005, Japan}}

\date{}

\maketitle

\begin{abstract}

Time-dependent processes accompanied by vortex creations and annihilations 
are investigated 
in terms of the eigenstates 
in conjugate spaces of Gel'fand triplets in 2-dimensions. 
Creations and annihilations of vortices are described 
by the insertions of unstable eigenstates with complex-energy eigenvalues 
into stable states written by the superposition of eigenstates 
with zero-energy eigenvalues. 
Some concrete examples are presented in terms of the eigenfunctions of 
the 2-dimensional parabolic potential barrier, i.e., $-m \gamma^2 (x^2+y^2)/2$. 
We show that the 
processes accompanied by vortex creations and annihilations can be analyzed 
in terms of the eigenfunctions in the conjugate spaces of Gel'fand triplets. 
Throughout these examinations 
we point out three interesting properties of the zero-energy flows. 
(i) Mechanisms 
using the zero-energy flows are absolutely economical from the viewpoint of energy 
consumption. 
(ii) An enormous amount of information
 can be discriminated in terms of the infinite topological 
variety of the zero-energy flows. 
(iii) The zero-energy flow patterns are absolutely stable in any disturbance 
by inserting arbitrary decaying flows with complex-energy eigenvalues. 

\vskip 5pt
Keywords: Zero-energy solutions, vortex creations and annihilations, 
quantum mechanics, Gel'fand triplets, 

\end{abstract}

\thispagestyle{empty}

\setcounter{page}{0}

\pagebreak

\section{Introduction} \label{sect.1.0}

Vortices play interesting roles in various aspects of 
present-day physics such as vortex matters (vortex lattices) 
in condensed matters~\cite{blat,crab}, quantum Hall effects
 [3-7], 
various vortex patterns of 
non-neutral plasma [8-11] 
and Bose-Einstein gases [12-16]. 
In hydrodynamics the vortices are well known objects and 
has been investigated in many aspects [17-21]. 
A hydrodynamical approach was also vigorously 
investigated in the early stage of the development 
of quantum mechanics [22-29], and 
some fundamental properties of vortices in quantum mechanics 
were extensively examined by many authors [30-37]. 

Recently we have proposed a way to investigate vortex patterns 
in terms of zero-energy solutions of Schr\"{o}dinger 
equations in 2-dimensions, which are eigenfunctions 
in conjugate spaces of Gel'fand triplets (CSGT) and 
degenerate infinitely~\cite{k1,ks8}. 
It should be noted that the eigenfunctions in CSGT represent 
scattering states, and thus they are generally not normalizable [40]. 
Therefore, the probability density ($|\psi|^2$) and 
the probability current 
(${\vj}={\rm Re}[\psi^*(-i\hbar \nabla)\psi]/m$) for the 
eigenfunction ($\psi$), which are defined  
in usual Hilbert spaces, cannot be introduced to the eigenfunctions 
in CSGT. 
Instead of the probability current, however, the velocity 
which is defined by ${\vv}={\vj}/|\psi|^2$ can have a 
well-defined meaning, because the ambiguity due to the 
normalization of the eigenfunctions disappears in the definition 
of the velocity. 
Actually we have shown that many interesting objects used in hydrodynamics 
such as the complex velocity potential can be introduced 
in the 2-dimensions of CSGT [41]. 
We can expect that the hydrodynamical approach is a quite hopeful 
framework in the investigation of phenomena described in CSGT. 
One should pay attention to 
two important facts obtained in the early works [38,39]. 
One is the fact that the zero-energy solutions are common 
over the two-dimensional central potentials such that  
$V_a(\rho)=-a^2g_a\rho^{2(a-1)}$ 
with $\rho=\sqrt{x^2+y^2}$ except $a=0$ 
and then similar vortex patters described by the zero-energy solutions 
appear in all such potentials. 
Actually zero-energy solutions for a definite number of $a$ 
can be 
transformed to solutions for arbitrary number of $a$ by conformal 
mappings~\cite{k1,ks8}. 
The other is that the zero-energy solutions are infinitely degenerate, 
and thus all energy eigenvalues in CSGT with the potentials 
$V_a(\rho)$ are infinitely degenerate because the 
addition of the infinitely degenerate zero-energy solutions 
to arbitrary eigenfunctions does not 
change the energy eigenvalues at all. 
It should also be pointed out that non-zero energy solutions in CSGT 
generally have complex energy eigenvalues like ${\cal E}=E \mp i \Gamma$ 
with $E$, $\Gamma \in \real$  
and then their time developments 
are described by the factors $e^{-(i E/\hbar \pm \Gamma/\hbar)t}$~\cite{bohm}. 
We see that stationary flows and time-dependent flows 
are, respectively, represented by the zero-energy solutions and the non-zero 
energy solutions in CSGT. 
From these facts 
it can be expected that in CSGT time-dependent flows 
in real processes are described by 
linear combinations of some stationary flows 
written by the zero-energy 
solutions and non-stationary flows (time-dependent flows) 
given by non-zero (complex) energy solutions. 
In such flows vortex patterns will become important 
objects to identify the situations of the flows. 
We can image two different types of the vortex patterns. 
One is the stationary vortex patterns,  
and the other is the vortex patterns changing in the time developments, which have 
been observed in real processes associated with vortices [8-16]. 
As presented in our early works ~\cite{k1,ks8}, the stationary vortex pattens can be 
described by the superposition of the zero-energy solutions, 
while the time-dependent ones will be done by putting the non-zero energy solutions 
into the superposition. 
In this paper we shall study time-dependent vortex patterns 
accompanied by vortex creations and annihilations 
in terms of the eigenfunctions of  2D-PPB in CSGT, 
since the eigenfunctions with non-zero energy (complex energy) solutions 
in CSGT are known only in the case of the PPB.
We shall also see that this approach can be one possibility to analyze 
various time-dependent vortex patterns in a rigorous framework of quantum mechanics 
in CSGT.

A brief review on the eigenfunctions of 2D-PPB [41] and also the vortices in the 
hydrodynamical approach of quantum mechanics is presented in the section 2. 
In the section 3 various vortex patterns accompanied by vortex creations 
and annihilations are investigated in terms of the 
eigenfunctions of the PPB given in the section 2. 
In the section 4 discussions on problems in this 
approach are performed  and an application of this model 
to processes transmitting an enormous 
amount of informarion by a very small energy consumption 
and also preserving the information (making memories) stably as well 
 is noticed.

\section{Brief review on the eigenfunctions of 2D-PPB and vortices} \label{sect.2.0}
{\bf 2-1 Eigenfunctions of 2D-PPB}

The infinitely degenerate eigenfunctions with the zero-energy for the potentials 
$V_a(\rho)$ 
have explicitly been given in the papers ~\cite{k1,ks8}, whereas the eigenfunctions 
with non-zero energies are known only in the case of  parabolic potential barriers (PPB)~
[41-47]. 
In order to make some concrete examples of time-dependent vortex patterns  
we perform the following considerations in terms of 
eigenfunctions of a two-dimensional parabolic potential barrier (2D-PPB) 
$V=-m\gamma^2(x^2+y^2)/2$,  
which are already obtained in the paper~\cite{sk4}.  
Here we shall summarize the results for the 2D-PPB, which will be used in the following 
discussions. 
It is trivial that the eigenfunctions in the 2D-PPB are represented by the multiples of 
those of the 1D-PPB as the eigenfunctions of 2D-harmonic oscillator (2D-HO) 
are done by those of the 1D-HO. 
The eigenfunctions of the 1D-PPB for $x$, which have pure imaginary energy eigenvalues 
$ \mp i( n_x+{1 \over 2})\hbar \gamma$, are given by 
\begin{equation}
u^{\pm}_{n_x}(x)=e^{\pm i\beta^2x^2/2}H^{\pm}_{n_x}(\beta x)\ \ \ \ 
(\beta \equiv \sqrt{m\gamma/\hbar}),
\end{equation} 
where $ H^{\pm}_{n_x}(\beta x)$ are the polynomials of degree $n_x$ written 
in terms of Hermite polynomials 
$H_{n}(\xi)$ with $\xi =\beta x$ as~\cite{sk,s2} 
\begin{equation}
H_{n}^\pm(\xi)=e^{\pm i n\pi/4} H_{n}(e^{\mp i\pi/4}\xi). 
\end{equation}
We have four different types of the eigenfunctions in the 2D-PPB~\cite{sk4}. 
Two of them 
$$U^{\pm\pm}_{n_xn_y}(x,y)\equiv u^{\pm}_{n_x}(x)u^{\pm}_{n_y}(y)$$  
with the energy eigenvalues ${\cal E}^{\pm\pm}_{n_xn_y}=\mp i(n_x+n_y+1)$, 
respectively,  represent diverging and converging flows. 
(See figs.~\ref{fig:1.1} and~\ref{fig:1.2}.) 
Some examples for the low degrees are obtained as follows; 
\begin{align}
U^{\pm\pm}_{00}(x,y)&=e^{\pm i\beta^2(x^2+y^2)/2} ,  \nonumber \\
U^{\pm\pm}_{10}(x,y)&=2\beta x e^{\pm i\beta^2(x^2+y^2)/2} , \nonumber \\ 
U^{\pm\pm}_{01}(x,y)&=2\beta y e^{\pm i\beta^2(x^2+y^2)/2} ,\nonumber \\
U^{\pm\pm}_{20}(x,y)&=(4\beta^2x^2 \mp 2i) e^{\pm i\beta^2(x^2+y^2)/2}, \nonumber \\
U^{\pm\pm}_{11}(x,y)&=4\beta^2xy e^{\pm i\beta^2(x^2+y^2)/2} , \nonumber \\
U^{\pm\pm}_{02}(x,y)&=(4\beta^2y^2 \mp2i) e^{\pm i\beta^2(x^2+y^2)/2} .
\end{align}
It is transparent that the eigenfunctions of angular momentums are constructed 
in terms of the linear combinations of these diverging and converging flows. 
Examples for some low angular momentums are presented in ref.~\cite{sk4}. 
The other two  
$$U^{\pm\mp}_{n_xn_y}(x,y)\equiv u^{\pm}_{n_x}(x)u^{\mp}_{n_y}(y)\ \ {\rm with}\ \  
{\cal E}^{\pm\mp}_{n_xn_y}(x,y)=\mp i(n_x-n_y)$$ 
are corner flows round the center. 
(See figs.~\ref{fig:1.3} and~\ref{fig:1.4}.) 
The zero-energy solutions that are common in the potentials $V_a(\rho)$ 
appear when $n_x=n_y$ is satisfied. 
A few examples of the zero-energy eigenfunctions are obtained as follows; 
\begin{align}
U^{\pm\mp}_{00}(x,y)&=e^{\pm i\beta^2(x^2-y^2)/2} ,  \nonumber \\
U^{\pm\mp}_{11}(x,y)&=4\beta^2 xy e^{\pm i\beta^2(x^2-y^2)/2} , \nonumber \\
U^{\pm\mp}_{22}(x,y)&=4[4\beta^4x^2 y^2+1\pm 2i\beta^2(x^2-y^2)]
              e^{\pm i\beta^2(x^2-y^2)/2}.
\end{align}
The existence of the infinitely degenerate zero-energy solutions brings 
the infinitely degeneracy to all the eigenstates with definite energies. 
This fact means that the energy and the other quantum numbers which are related to 
the determination of the energy eigenvalues like angular momentums 
are not enough to discriminate the eigenstates. 
What are the quantum numbers to characterize the infinite degeneracy? 
One interesting candidate to characterize the states is vortex patterns, 
which can be observable in experiments [8-16]. 
Time-dependences of the patterns are also good candidates for observables 
in those processes. 
It is obvious that the eigenfunctions of (13) and (14) are not normalizable. 
The details on the proof that they are the eigenfunctions of CSGT are presented 
in refs. [46,47]. 
\vskip1pt 
\hfil\break
{\bf 2-2 Vortices in quantum mechanics}

Let us here briefly note 
how vortices are interpreted in quantum mechanics. 
The probability density $\rho(t,x,y)$ and 
the probability current $\vj(t,x,y)$ of a wavefunction $\psi(t,x,y)$ 
in non-relativistic quantum mechanics 
are defined 
by 
\begin{align}
  \rho(t,x,y)&\equiv\left| \psi(t,x,y)\right|^2, \\
  \vj(t,x,y)&\equiv{\rm Re}\left[\psi(t,x,y)^*
  \left(-i\hslash\nabla\right)\psi(t,x,y)\right]/m.
\end{align}
They satisfy the equation of continuity 
 $ \partial\rho/\partial t+\nabla\cdot\vj=0.$ 
Following the analogue of the hydrodynamical 
approach [17-21], 
the fluid 
can be represented by 
the density $\rho$ and the fluid velocity $\vv$. 
They satisfy Euler's equation of continuity 
\begin{equation}
  \frac{\partial\rho}{\partial t}+\nabla\cdot(\rho\vv)=0. 
\end{equation} 
Comparing this equation with the continuity equation, 
the following 
definition for the quantum velocity of the state $\psi(t,x,y)$ 
is led in the hydrodynamical approach; 
\begin{equation}
  \vv\equiv\frac{\vj(t,x,y)}{\left| \psi(t,x,y)\right|^2}. 
\end{equation}
Now it is obvious that vortices appear at the zero points of 
the density, that is, the nodal points of the wavefunction.  
At the vortices, of course, 
the current $\vj$ must not vanish. 
When we write the wavefunction $\psi(t,x,y)=\sqrt{\rho (x,y)} e^{iS(x,y)/\hbar }$, 
the velocity is given by ${\vv}=\nabla S/m$. 
We should here remember that the solutions 
degenerate infinitely. 
This fact indicates that we can construct wavefunctions having 
the nodal points at arbitrary positions in terms of 
linear combinations of the infinitely degenerate 
solutions~\cite{k1,ks8,sk4}. 

The strength of vortex is characterized 
by the circulation $\Gammav$ 
that is represented by the integral round a closed contour $C$
encircling the vortex such that 
\begin{equation}
\Gammav=\oint_C \vv \cdot d\vs
\end{equation} 
and it is quantized as 
\begin{equation}
\Gammav=2\pi l\hbar/m, 
\end{equation} 
where the circulation number $l$ is 
an integer~\cite{joh2,joh5,wu-sp,bb2}.
\vskip10pt

\section{Stationary vortex patterns and 
vortex creations and annihilations} \label{sect.3.0}

Let us investigate vortex patterns in terms of the 2D-PPB eigenfunctions 
given in (3) and (4).
Since the zero-energy solution given in (4) have no nodal point, 
they have no vortex. 
However, it has been shown that 
some vortex patterns having infinite numbers of vortices, 
like vortex lines 
and vortex lattices, can be made in terms of simple linear combinations of 
those low lying stationary states in the early works~\cite{k1,ks8}. 
We can, of course, make linear combinations having a few or some vortices. 
In the following discussions we shall study compositions 
not only for stationary vortex patterns but also for 
patterns accompanied by vortex creations and 
annihilations on the basis of a stationary flow having no vortex
(we shall call it the basic flow hereafter). 
In these composition processes we shall see that 
various vortex patterns will be an understandable problem
in quantum mechanics by considering eigenstates in CSGT. 
As the basic flow without any vortices we take the flow described by 
\begin{align} 
\Phi_B&={1 \over 4} U_{22}^{+-}(x,y) -U_{00}^{+-}(x,y)    \nonumber \\
      &=(4\beta^4x^2y^2+2i\beta^2(x^2-y^2))e^{i\beta^2(x^2-y^2)/2}. 
\end{align} 
Note that the nodal point at the origin does not produce vortices, 
because the current vanishes there. 

\hfil\break
{\bf 3-1 Stationary vortex patterns}

At first we shall show two 
stationary patterns with three and four vortices, which will be 
used in the following discussions on vortex creations and  annihilations.
\hfil\break 
{\bf (1-1) Three vortices pattern}

The linear combination of the basic flow 
and $U^{+-}_{22}$ such that 
\begin{align}
\Phi^{+-}_{012}(x,y)&=\Phi_B-c^2 U^{+-}_{11}(x,y) \nonumber \\
             &=[4\beta^2xy(\beta^2xy-c^2)+2i\beta^2(x^2-y^2)]
             e^{i\beta^2(x^2-y^2)/2},
\end{align}
with $c \in \real$ 
has three vortices at the origin and the two points $( \pm c/\beta, \pm c/\beta)$ 
as shown in fig.~\ref{fig:1.5}. 
If we take $-c^2$ instead of $c^2$, three vortices appear at the origin and the points 
$( \pm c/\beta, \mp c/\beta)$.
\hfil\break
{\bf (1-2) Four vortices pattern }

The linear combinations given by 
\begin{align}
\Phi^{+-}_{02}(x,y)&={1 \over 4}\Phi_B-c^4U^{+-}_{00}(x,y) \nonumber \\
             &=[(\beta^2xy-c^2)(\beta^2xy+c^2)+i\beta^2(x^2-y^2)/2]
             e^{i\beta^2(x^2-y^2)/2},
\end{align} 
has four vortices 
at the points $( \pm c/\beta, \pm c/\beta)$ and 
$( \pm c/\beta, \mp 1c/\beta)$ as shown in fig.~\ref{fig:1.6}. 

\hfil\break
{\bf 3-2 Time-dependent patterns accompanied by vortex creations and annihilations}

Let us go to the study of time-dependent vortex patterns 
that are obtained by linear combinations of 
stationary flows and time dependent ones. 
We shall see some simple vortex creation and annihilation processes. 
In the following considerations the time dependent flows are put in 
the stationary flows at $t=0$. 

\hfil\break
Let us first discuss the processes 
where some vortices are once created and then they are annihilated 
in the time development. 
\hfil\break
{\bf (2-1) A pattern accompanied by a pair creation and annihilation}

Let us consider the following linear combination; 
\begin{align}
\Phi^{+-}_{02,1}(x,y,t)&={1 \over 2}\Phi_B-\theta(t)c^3U^{+-}_{10}(x,y)
          e^{-\gamma t} \nonumber \\
             &=[2\beta x(\beta^3xy^2-\theta(t)c^3e^{-\gamma t})+i\beta^2(x^2-y^2)]
          e^{i\beta^2(x^2-y^2)/2}, 
\end{align} 
where the theta function is taken as 
$\theta (t)=0$ for $t<0$ and $=1$ for $t\geq 0$. 
It has two nodal points at 
$(c e^{-\gamma t/3}/\beta,\pm c e^{-\gamma t/3}/\beta)$ for $t\geq 0$, 
where two vortices with opposite circulation numbers  exist. 
The nodal points go to the origin as the time $t$ goes to infinity as shown 
in fig.~\ref{fig:1.7}. 
Since the contribution of the unstable flow 
decreases 
as $t \rightarrow \infty $ because of the time factor $e^{-\gamma t}$,  
the wavefunction 
$\Phi^{+-}_{02,1}(x,y,t)$ 
goes to $\Phi_B$ as $t \rightarrow \infty$. 
Thus the flow has no nodal point in the limit. 
This means that the pair of vortices which are created at $t=0$ 
disappears at origin in the limit $t \rightarrow \infty $. 
We can say that this wavefunction describes the pair annihilation of two vortices. 
The time development of this process can be described as follows: 
\hfil\break 
(i) Before the time-dependent flow is put in the basic flow, i.e., $t<0$, 
there is no vortex. 
\hfil\break 
(ii) At $t=0$ when the time-dependent flow is put in the basic flow, 
a pair of vortices are suddenly created. 
\hfil\break 
(iii) The pair moves toward the origin, and then they annihilate at the origin, 
that is, the flow turns back to the basic flow $\Phi_B$ having no vortex. 
\hfil\break
{\bf (2-2) A pattern accompanied by creations and annihilations of two pairs}

The linear combination given by 
\begin{align}
\Phi^{+-}_{02,2}(x,y,t)&=\Phi_B-\theta(t)c^2[2iU^{+-}_{00}(x,y)
                -U^{+-}_{20}(x,y)e^{-2\gamma t}] \nonumber \\
  &=[4\beta^2x^2(\beta^2y^2-\theta(t)c^2e^{-2\gamma t})-
         2i(\theta(t)c^2(1-e^{-2\gamma t})-\beta^2(x^2-y^2))]
             e^{i\beta^2(x^2-y^2)/2}
\end{align} 
has four nodal points at $(\pm c/\beta, c e^{-\gamma t}/\beta)$ and 
$(\pm c/\beta, -c e^{-\gamma t}/\beta)$ for $t\geq 0$. 
In this case we easily see that 
the pair of vortices at $(\pm c/\beta, c e^{-\gamma t}/\beta)$ and that 
at $(\pm c/\beta, -c e^{-\gamma t}/\beta)$ annihilate as $t \rightarrow \infty $ 
as shown in fig.~\ref{fig:1.8}.  
The stationary flow $\Phi_B-2ic^2U^{+-}_{00}(x,y)$ that appears in the limit 
has two nodal points at $(\pm c/\beta,0)$, but it has no vortex, because the 
current also vanish at the points. 
The time development of this process is interpreted similarly as the case (1-1). 

In these two cases all vortices move on straight lines in the pair annihilation 
processes.
\hfil\break
{\bf (2-3) A pattern accompanied by creation and annihilation of four vortices}

The linear combinations of stationary solutions and diverging or converging 
flows make different types of annihilation processes. 
For an example, let us consider the linear combination of $ \Phi_B$ and 
the lowest order diverging flow described by 
\begin{equation}
U_{00}^{++}(x,y,t)=e^{i\beta^2(x^2+y^2)/2}e^{-\gamma t}
\end{equation}  
having the energy eigenvalue $-i \gamma \hbar$. 
Let us consider the linear combination described by 
\begin{align}
\Phi^{++}_{02,0}(x,y,t)&={1 \over 4}\Phi_B
                -\theta(t)c^2 U^{++}_{00}(x,y,t) \nonumber \\
  &=[\beta^4x^2y^2-\theta(t)c^2 e^{-\gamma t}e^{i\beta^2y^2}+i\beta^2(x^2-y^2)/2]
             e^{i\beta^2(x^2-y^2)/2}.
\end{align} 
For $t\geq 0$ it has two nodal points at the points where the following relations 
are fulfilled; 
\begin{equation}
XY=c(t) {\rm cos}Y,\ \ \ {1 \over 2}(X-Y)-c(t){\rm sin}Y=0, 
\end{equation} 
where $X=\beta^2 x^2$, $Y=\beta^2 y^2$ and $c(t)=c^2 e^{-\gamma t}$. 
From these relations we have an equation for $Y$  
\begin{equation}
Y^2-c(t) {\rm cos}Y +2c(t)Y{\rm sin}Y=0. 
\end{equation} 
The solutions are obtained from the cross points 
of two functions 
$f(Y)=Y^2$ and $g(Y)=c(t)({\rm cos}Y-2Y{\rm sin}Y)$. 
We easily see that a solution for $Y\geq 0$ exists in the region 
$0<Y<\pi /2$ for arbitrary positive numbers of $c(t)$. 
Four vortices appear at the four points expressed by the combinations of 
$x=\pm \sqrt{X}/\beta $ and $y=\pm \sqrt{Y}/\beta$, where 
$X$ is obtained by using the first relation of (18). 
Since $c(t)$ goes to $0$ as $t \rightarrow \infty $, we see that 
 $X$ and $Y$ 
simultaneously go to $0$ in the limit such that 
$$
X\simeq Y\rightarrow |c|e^{-\gamma t/2} \rightarrow 0, \ \ \ 
{\rm for}\ \  t \rightarrow \infty .  
$$ 
Since the flow turns back to the basic flow, the four vortices 
annihilate at the origin in the limit. 
From the second relation of (18), 
we have 
$$
X=Y+2c(t) {\rm sin}Y.
$$ 
This equation show us that the vortex points do not 
move toward the origin along straight lines. 

\hfil\break
Here we consider two somewhat complicated processes. 
\hfil\break
{\bf (2-4) A pattern accompanied by a pair annihilation and creation}

 Let us start from the stationary flow described by the wavefunction 
$\Phi^{+-}_{012}(x,y)$ of (12), which has three vortices. 
We consider a linear combination of the stationary state and 
a time-dependent flow described by
\begin{equation}
\Psi^{+-}_{2}(x,y;t)= (U^{+-}_{20}+U^{+-}_{02})e^{-2\gamma t}
\end{equation} 
such that 
\begin{align}
\Phi^{+-}_{012,2}(x,y;t)&=\Phi^{+-}_{012}(x,y)+
        \theta(t){1 \over 2}ai\Psi^{+-}_{2}(x,y;t) \nonumber \\
     &=\left\{4\beta^2xy(\beta^2xy-c^2)+2i[(1+\theta(t)a(t))\beta^2x^2-
     (1-\theta(t)a(t))\beta^2y^2]
        \right\}e^{i\beta^2 (x^2-y^2)/2},
\end{align} 
where $a\in \real$ and $a(t)=ae^{-2\gamma t}$. 
For $t\geq 0$ 
the nodal points of the wavefunction appear at the points fulfilled 
the following relations;
\begin{equation}
\xi\eta(\xi\eta-c^2)=0,   \ \ \ \ (1+a(t))\xi^2-(1-a(t))\eta^2=0,
\end{equation} 
where $\xi=\beta x$ and $\eta=\beta y$. 
We easily see that for 
$
|a(t)|>1
$ 
the second relation has only one solution at $\xi=\eta=0$, 
whereas for $|a(t)|<1$ it has two non-zero solutions at 
\begin{equation} 
\xi(t)=\pm({1-a(t) \over 1+a(t)})^{1/4}c, 
\ \ \ \eta(t)=\pm ({1+a(t) \over 1-a(t)})^{1/4}c.
\end{equation}   
This means that at the critical time $t_c$ 
satisfying the relation $|a(t_c)|=1$, that is, 
\begin{equation}
t_c={1 \over \gamma } {\rm ln}|a| 
\end{equation} 
a pair of vortices are suddenly created at the two points 
($\xi(t_c)/\beta , \eta(t_c)/\beta $). 
We can interpret this process for $|a|>1$ as follows 
(see also figs.~\ref{fig:1.9},~\ref{fig:1.10},~\ref{fig:1.11} and~\ref{fig:1.12}): 
\hfil\break
(i) Before the shock described by the time-dependent flow $\Psi^{+-}_{2}(x,y;t)$ 
is given, the flow is described by the stationary flow $\Phi^{+-}_{012}(x,y)$ 
with three vortices given in fig. 9.
\hfil\break
(ii) At $t=0$ the shock is given, and then two vortices at two points 
$(\pm c/\beta ,\pm c/\beta)$ suddenly disappear. (See fig. 10.)
\hfil\break
(iii) At the critical time $t=t_c$ two vortices again suddenly appear at 
($\xi(t_c)/\beta , \eta(t_c)/\beta $), and they move toward the original positions 
$(\pm c/\beta,\pm c/\beta)$ for $t>t_c$. (See fig. 11.)
\hfil\break
(iv) Finally the flow turns back to the original stationary flow. (See fig. 12.)
\hfil\break
Throughout this process the vortex being at the origin does not move at all.
\hfil\break
{\bf (2-5) A pattern accompanied by eight vortices annihilation } 

Here we take $\Phi^{+-}_{02}(x,y)$ of (13) as the stationary flow, 
which has four vortices. 
For the simplicity $c=1/2$ is taken in the following discussions. 
Here the lowest order diverging flow 
$U_{00}^{++}(x,y,t)$ 
 are put into the stationary flow at $t=0$. 
The wavefunction are given by 
\begin{align}
\Phi_{02,0}^{++}(x,y;t)=&16\Phi^{+-}_{02}(x,y)+\theta(t)b^2U_{00}^{++}(x,y,t)  \nonumber  \\
                       =&[16\beta^4x^2y^2-1+\theta(t)b^2 e^{-\gamma t}e^{i\beta^2y^2}+
                      8i\beta^2(x^2-y^2)]
             e^{i\beta^2(x^2-y^2)/2} \nonumber \\
             =&[16\beta^4x^2y^2-1+\theta(t)b(t) e^{-\gamma t}{\rm cos}(\beta^2y^2)
               +i(8\beta^2(x^2-y^2)   \nonumber \\
             &+
             \theta(t)b(t) e^{-\gamma t}{\rm sin}(\beta^2y^2))]
             e^{i\beta^2(x^2-y^2)/2},
\end{align} 
where $b\in \real $ and $b(t) =b^2e^{-\gamma t}$. 
Using $X=\beta^2x^2$ and $Y=\beta^2y^2$, we have two relations for nodal points 
of the wavefunction for $t>0$ as follows;
\begin{equation} 
16X Y +b(t) {\rm cos}Y-1=0, \ \ \ \ 
8(X-Y)+b(t){\rm sin}Y=0,
\end{equation} 
From these relations we obtain an equation for the nodal points 
\begin{equation} 
1-b(t){\rm cos}Y -16Y^2+2b(t)Y{\rm sin}Y=0.
\end{equation} 
Examining the cross point of the two functions 
$F(Y)=16Y^2-1$ and $G(Y)=-b(t)({\rm cos}Y -2Y{\rm sin}Y)$, 
we obtain the following results:   
\hfil\break
(1) In the case of $b(t)<1$ 
the two functions always have a cross-point in the region satisfying 
$Y\geq 0$ (note that $Y=\beta^2y^2$). 
The wavefunction, therefore, has four nodal points. 
This means that the flow always has four vortices that move toward the stationary 
points fulfilling $|x|=|y|=(2 \beta)^{-1}$ as $t$ increases. 
\hfil\break
(2) In the case of $b(t)>1$ eq.(27) has an even number of solutions like 
$n=0,2,4,\cdots$. 
Since one solution brings four vortices on a circle with the center at the origin, 
the vortex number is given by $4n$. 
Note that the number $n$ increases as $b(t)$ increases. 
This fact means that, since $b(t)$ decreases as $t$ increases, the vortex number 
decreases as $t$ increases until $b(t)$ gets to 1. 
Considering that the change of $n$ is always 2, we see that 
the reduction of the vortex number caused by the change of $n$ is always 8. 
That is to say, we observe that four vortex-pairs simultaneously annihilate 
at four different points on a circle. 
As a simple example, let us consider the case of $n=2$ at $t=0$. 
We observe the following time development of the flow: 
   \hfil\break
(i) 
For $t<0$ the stationary flow has the four vortices as shown in fig.6. 
   \hfil\break
(ii) 
At $t=0$ the original four vortices are disappear and 
eight vortices are newly created. 
Then we observe 
the flow having eight vortices. (See fig.~\ref{fig:1.13}.) 
    \hfil\break
(iii) 
In the time-development the eight vortices disappear simultaneously. 
We observe the process as the annihilations of four vortex-pairs. 
Thus the flow having no vortex appears. 
    \hfil\break
(iv) 
At the critical time $t_c= {\rm ln} b^2 /\gamma$ 
when $b(t_c)=1$ is fulfilled  four vortices are created at the origin, and then 
they move toward the stationary points. 
The vortex state at $t=t_c$ can be understood as a vortex quadrupole~\cite{k1}. 

If, instead of the diverging flow  $U_{00}^{++}(x,y,t)$, 
the converging flow $U_{00}^{--}(x,y,t)$ is put in the stationary flow, 
we observe a flow continuously creating 8 vortices for any choices of $b$. 
Of course, the time dependent flow blows up the magnitude in the limit 
of $t\rightarrow \infty$, and thus the original stationary flow can not be 
observed in the limit. 

\section{Discussions} \label{sect.4.0}

We have shown concrete examples of different types of 
vortex patterns accompanied by 
creations and annihilations of vortices by using only some low degree 
solutions of the 2D-PPB. 
We can, of course, present more complicated patterns 
by introducing the higher degree solutions, but the examples 
presented in the section 3 
will be enough to show the fact that various vortex patterns 
can be reproducible in terms of the eigenfunctions of the 2D-PPB. 
As already noted that the zero-energy solutions in the 2D-PPB 
can be transformed to those in the potentials $V_a(\rho )$ by the 
conformal mappings [38,39], the stationary patterns given in the section 3-1 
can be mapped into the stationary patterns of arbitrary potentials. 
This fact means that, as far as the stationary patterns are concerned, 
there is an exact one-to-one correspondence between the patterns of 
the PPB and those of the other potentials. 
As for the time-dependent patterns we cannot present any concrete examples 
except the case of the 2D-PPB at this moment, 
but we may expect that similar vortex patterns as those given in the section 
3-2 for the PPB will appear in other potentials, 
since all energy eigenstates with complex eigenvalues degenerate infinitely 
in all the potentials $V_a(\rho )$ as same as in the 2D-PPB. 
Anyhow we cannot exactly say about the problem before we find any solutions 
with complex eigenvalues in the other cases. 

Here we would like to note 2D-PPB. 
We do not know any physical phenomena that are described by 2D-PPB. 
We can, however, expect that most of weak repulsive forces in matters composed of 
many constituents will be approximated by PPBs as most of weak attractive forces 
are well approximated by harmonic oscillators. 
In general flows that go round a smooth hill of potential feel a weak 
repulsive force represented by a PPB [48-50]. 
Actually we see that when a charged particle is put in an infinitely 
long tube where same charged particles are uniformly distributed, 
the charged particle feels 2D-PPB. 
In non-neutral plasma  electrons being near the center will possibly be in 
a similar situation. 
In the plasma electro-magnetic interactions must be introduced. 
It should be noted that in the case of a charged particle in a magnetic field 
the vortex quantization given by (9) can be read as 
\begin{align}
 m \Gammav &=\oint _c (\nabla S -q{\bf A})\cdot d{\bf s} \nonumber \\
        &=\oint _c {\bf p}\cdot d{\bf s} - q\Phi ,
 \end{align} 
 where $q$ is the charge of the particle and 
 $\Phi$ is the magnetic flux passing through the enclosed surface. 
Analyzing vortex phenomena of non-neutral plasma in terms of the  
eigenfunctions of the 2D-PPB will be an interesting application.

The infinite freedom arising from the infinite degeneracy 
of the zero-energy solutions should be noticed. 
Such a freedom has never appeared in the statistical mechanics describing 
thermal equilibrium. 
The freedom is different from that generating the usual entropy and then 
temperatures, because the freedom does not change real energy 
observed in experiments at all. 
A model of statistical mechanics for the new freedom has been proposed 
and some simple applications have been performed in the case of 
1D-PBB [51-53]. 
The model is applicable to slowly changing phenomena in the time 
evolutions, because the PPB has only pure imaginary energy eigenvalues 
in the 1-dimension. 
In the present model of the 2-dimensions, however, 
we have the infinite degree of freedom 
arising from the zero-energy solutions that have no time evolution. 
The huge degeneracy of the zero-energy solutions can provide the huge 
variety in every energy eigenstates, which will be identified by 
the vortex patterns. 
In such a consideration the vortex patterns will be understood as 
the topological properties of flows. 
How this freedom should be counted in statistical mechanics is an important 
problem in future considerations. 

Finally we would like to comment that even in thermal equilibrium 
the freedom of the  zero-energy flows can be used freely.   
Furthermore the use of the zero-energy flows is very useful and economical 
from the viewpoint of energy consumption. 
For example it can be a very economical step  
for the transmission of information. 
The transmission by the use of the zero-energy flows 
having the huge variety 
enable us to transmit an enormous 
amount of information without any energy loss. 
Considering that the flows are stationary, 
they can also be a very useful step for making mechanisms to preserve 
such information, i.e., for memories in living beings. 
The huge variety of vortex patterns can possibly discriminate 
the enormous amount of information.  
The change of the preserved memories can easily be carried out by 
changing the vortex patterns by the insertion of some 
zero-energy (stationary) flows in the preserved ones 
(see the processes presented in the section 3-1). 
Furthermore, as shown in the section 3-2, 
in all the time-dependent processes 
induced by the insertion of the unstable flows 
with complex energy eigenvalues the time-dependent flows always turn back 
to the stationary flows in the long time scales, 
and then the initial flow patterns are recovered. 
That is to say, the initial patterns are kept in all such time-dependent processes. 
This stability of the flow patterns seems to be a very interesting property 
for the interpretation of the stability of memories not only in their preservations  
but in their applications as well. 
The applications, of course, mean thinking processes.     
The flows will possibly be workable in the steps for  
thinking in living beings. 
The use of the zero-energy and complex-energy solutions 
enables living beings to make up many functions in their bodies
very economically on the basis of energy consumption. 
Anyway the zero-energy and complex-energy solutions are 
interesting objects to describe 
mechanisms working very economically as for the energy consumption. 
Especially, in the 2D-PPB case we can do it without any energy loss, 
because all the solutions of PPB have no real energy eigenvalue, 
i.e., the energy eigenvalues are zero or pure imaginary. 
Here we would like to summarize the property of flows in CSGT. 
As for the zero-energy flows we should stress the following three 
properties; 
(i) the absolutely energy-saving mechanism, 
(ii) the mechanism including an enormous topological variety, and 
(iii) the perfect recovery of the flow patterns in any disturbance by 
the insertions of arbitrary decaying flows with complex-energy eigenvalues. 
On the other hand the role of the flows with complex energies can be 
understood as the excitement mechanism among the zero-energy flows. 
We still have a lot of problems to overcome the present situation, but 
we may expect that 
the study of the zero- and complex-energy solutions in CSGT will open 
a new site in physics.


\pagebreak

  \begin{figure}
   \begin{center}
    \begin{picture}(200,200)
     \thicklines
     \put(0,100){\vector(1,0){200}}
     \put(100,0){\vector(0,1){200}}
     \put(90,88){$0$}
     \put(205,98){$x$}
     \put(98,205){$y$}
     
     \put(105,105){\vector(1,1){90}}
     \put(95,105){\vector(-1,1){90}}
     \put(85,85){\vector(-1,-1){80}}
     \put(105,95){\vector(1,-1){90}}
    \end{picture}
   \end{center}
   \caption[]{Diverging flows.}
   \label{fig:1.1}
  \end{figure}
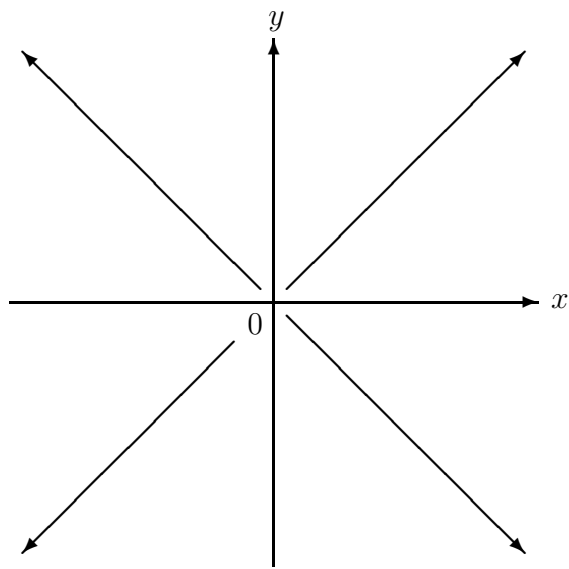

  \begin{figure}
   \begin{center}
    \begin{picture}(200,200)
     \thicklines
     \put(0,100){\vector(1,0){200}}
     \put(100,0){\vector(0,1){200}}
     \put(90,88){$0$}
     \put(205,98){$x$}
     \put(98,205){$y$}
     
     \put(195,195){\vector(-1,-1){90}}
     \put(5,195){\vector(1,-1){90}}
     \put(5,5){\vector(1,1){80}}
     \put(195,5){\vector(-1,1){90}}
    \end{picture}
   \end{center}
   \caption[]{Converging flows.}
   \label{fig:1.2}
  \end{figure}
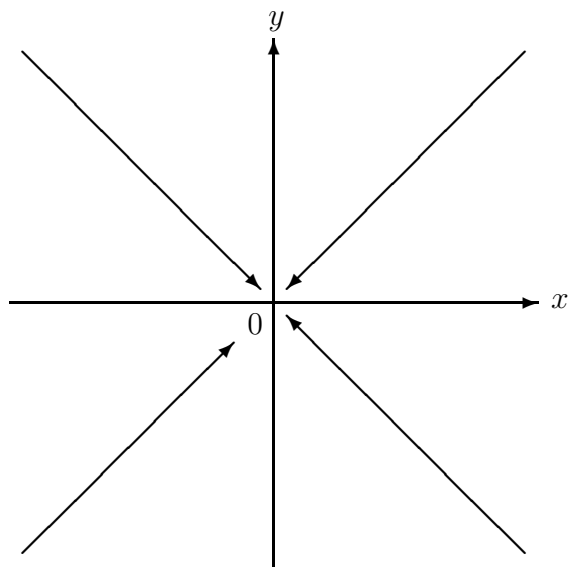

  \begin{figure}[pb]
   \begin{center}
    \begin{picture}(200,200)
     \thicklines
     \put(0,100){\vector(1,0){200}}
     \put(100,0){\vector(0,1){200}}
     \put(90,88){$0$}
     \put(205,98){$x$}
     \put(98,205){$y$}
     
     \qbezier(105,200)(105,105)(200,105)
     \qbezier(95,200)(95,105)(0,105)
     \qbezier(95,0)(95,95)(0,95)
     \qbezier(105,0)(105,95)(200,95)
     
     \put(200,105){\vector(1,0){1}}
     \put(0,105){\vector(-1,0){1}}
     \put(0,95){\vector(-1,0){1}}
     \put(200,95){\vector(1,0){1}}
    \end{picture}
   \end{center}
   \caption[]{Corner flows moving from the $y$-direction 
   to the $x$-direction.}
   \label{fig:1.3}
  \end{figure}
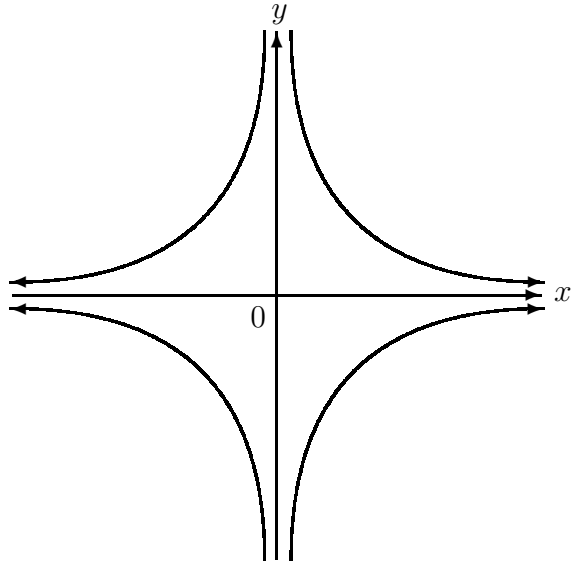

  \begin{figure}
   \begin{center}
    \begin{picture}(200,200)
     \thicklines
     \put(0,100){\vector(1,0){200}}
     \put(100,0){\vector(0,1){200}}
     \put(90,88){$0$}
     \put(205,98){$x$}
     \put(98,205){$y$}
     
     \qbezier(105,200)(105,105)(200,105)
     \qbezier(95,200)(95,105)(0,105)
     \qbezier(95,0)(95,95)(0,95)
     \qbezier(105,0)(105,95)(200,95)
     
     \put(105.5,200){\vector(0,1){1}}
     \put(95.5,200){\vector(0,1){1}}
     \put(95.5,0){\vector(0,-1){1}}
     \put(105.5,0){\vector(0,-1){1}}
    \end{picture}
   \end{center}
   \caption[]{Corner flows moving from the $x$-direction 
   to the $y$-direction.}
   \label{fig:1.4}
   \end{figure}
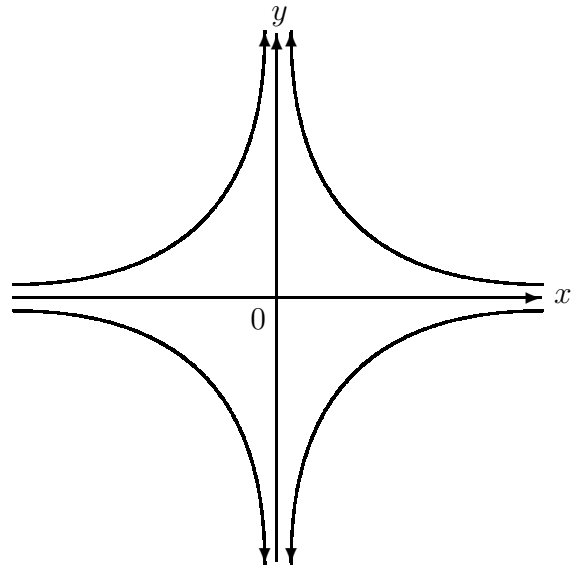
  
   
    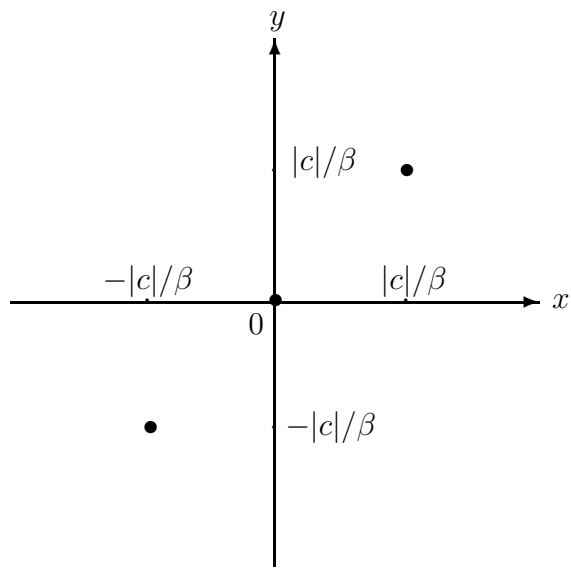
\begin{figure}
    \begin{center}
    \begin{picture}(200,200)
     \thicklines
     \put(0,100){\vector(1,0){200}}
     \put(100,0){\vector(0,1){200}}
     \put(90,88){$0$}
     \put(205,98){$x$}
     \put(98,205){$y$}
     
     \put(148,98){$\cdot $}
     \put(50,98){$\cdot $}
     \put(140,105){$|c|/\beta $}
     \put(35,105){$-|c|/\beta $}
     \put(98,147){$\cdot $}
     \put(98,50){$\cdot $}
     \put(106,150){$|c|/\beta $}
     \put(104,50){$-|c|/\beta $}

     \put(147,147){$\bullet$}
     \put(97.4,97.8){$\bullet$ }
     \put(50,50){$\bullet $}
    \end{picture}
   \end{center}
   \caption[]{Stationary pattern with three vortices.
        $\bullet$ denotes a vortex.}
   \label{fig:1.5}
    \end{figure}

    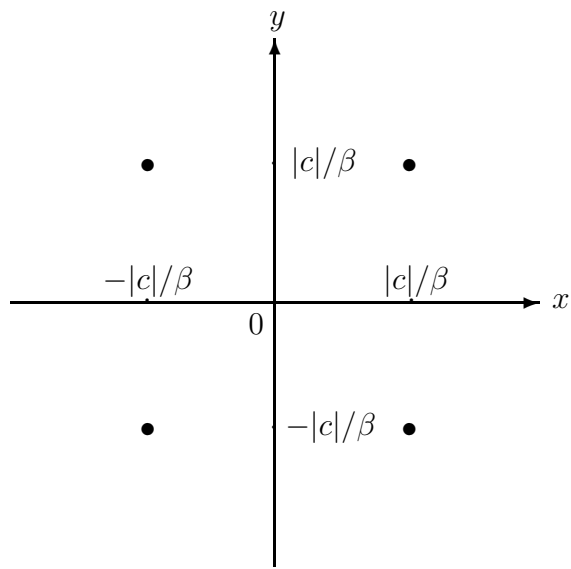
\begin{figure}
    \begin{center}
    \begin{picture}(200,200)
     \thicklines
     \put(0,100){\vector(1,0){200}}
     \put(100,0){\vector(0,1){200}}
     \put(90,88){$0$}
     \put(205,98){$x$}
     \put(98,205){$y$}
     
     \put(150,98){$\cdot $}
     \put(50,98){$\cdot $}
     \put(141,105){$|c|/\beta $}
     \put(35,105){$-|c|/\beta $}
     \put(98,150){$\cdot $}
     \put(98,50){$\cdot $}
     \put(106,150){$|c|/\beta $}
     \put(104,50){$-|c|/\beta $}

     \put(148,149){$\bullet $}
     \put(148,49){$\bullet $}
     \put(49,149){$\bullet $}
     \put(49,49){$\bullet $}
    \end{picture}
   \end{center}
   \caption[]{Stationary pattern with four vortices.}
   \label{fig:1.6}
    \end{figure}
   
   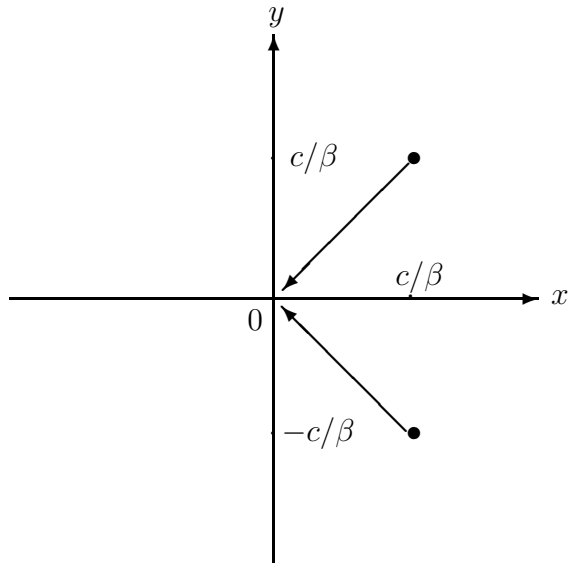
\begin{figure}
    \begin{center}
    \begin{picture}(200,200)
     \thicklines
     \put(0,100){\vector(1,0){200}}
     \put(100,0){\vector(0,1){200}}
     \put(90,88){$0$}
     \put(205,98){$x$}
     \put(98,205){$y$}
     
      \put(150,150){$\bullet$}
     \put(150,46){$\bullet $}
     
     \put(150,98){$\cdot $}
     \put(146,105){$c/\beta $}
     \put(98,150){$\cdot $}
     \put(98,46){$\cdot $}
     \put(106,150){$c/\beta $}
     \put(103,46){$-c/\beta $}
     
     \put(151,151){\vector(-1,-1){47.65}}
     \put(150,50){\vector(-1,1){47}}
    \end{picture}
   \end{center}
   \caption[]{A pair annihilation pattern for $c>0$. 
   The arrows show the moving directions of the vortices.}
   \label{fig:1.7}
  \end{figure}

  \begin{figure}
   \begin{center}
    \begin{picture}(200,200)
     \thicklines
     \put(0,100){\vector(1,0){200}}
     \put(100,0){\vector(0,1){200}}
     \put(90,88){$0$}
     \put(205,98){$x$}
     \put(98,205){$y$}
     
      \put(150,150){$\bullet $}
     \put(150,50){$\bullet $}
     \put(48,150){$\bullet $}
     \put(48,50){$\bullet $}

     \put(151,98){$\cdot $}
     \put(48.5,98){$\cdot $}
     \put(156,105){$|c|/\beta $}
     \put(14,105){$-|c|/\beta $}
     \put(98,150){$\cdot $}
     \put(98,50){$\cdot $}
     \put(106,150){$|c|/\beta $}
     \put(103,50){$-|c|/\beta $}

     \put(152.4,150){\vector(0,-1){47}}
     \put(152.5,52){\vector(0,1){46}}
     \put(50.4,52){\vector(0,1){46}}
     \put(50.4,150){\vector(0,-1){47}}
    \end{picture}
   \end{center}
   \caption[]{Two pairs annihilation pattern.}
   \label{fig:1.8}
  \end{figure}
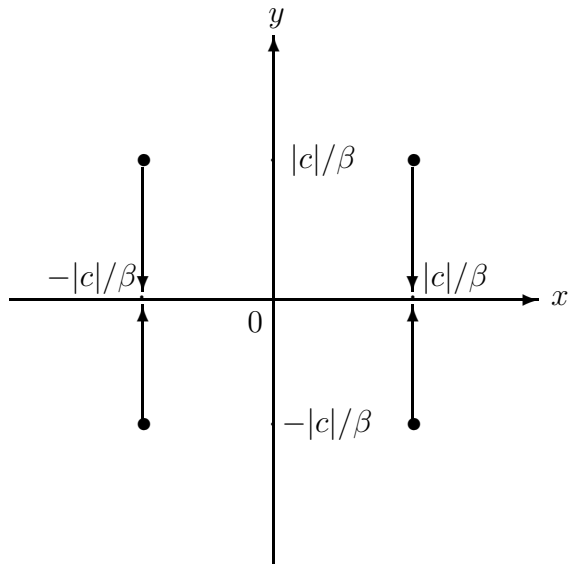

    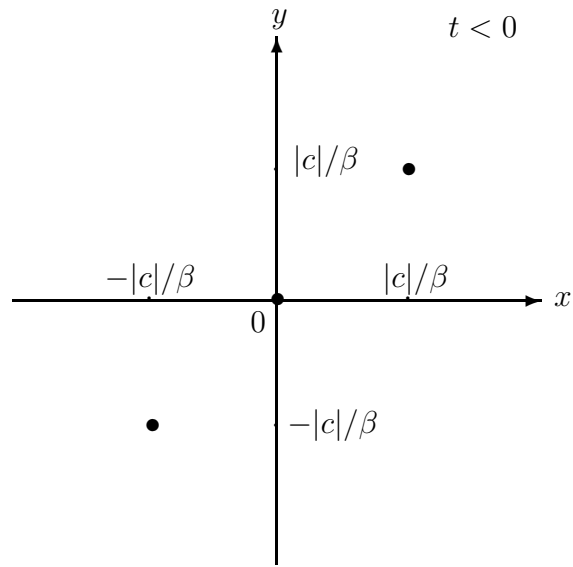
\begin{figure}
    \begin{center}
    \begin{picture}(200,200)
     \thicklines
     \put(0,100){\vector(1,0){200}}
     \put(100,0){\vector(0,1){200}}
     \put(90,88){$0$}
     \put(205,98){$x$}
     \put(98,205){$y$}
     \put(165,200){$t<0$}
     
     \put(148,98){$\cdot $}
     \put(50,98){$\cdot $}
     \put(140,105){$|c|/\beta $}
     \put(35,105){$-|c|/\beta $}
     \put(98,147){$\cdot $}
     \put(98,50){$\cdot $}
     \put(106,150){$|c|/\beta $}
     \put(104,50){$-|c|/\beta $}

     \put(147,147){$\bullet$}
     \put(97.4,97.8){$\bullet$ }
     \put(50,50){$\bullet $}
    \end{picture}
   \end{center}
   \caption[]{Initial vortex pattern for $t<0$.
       }
   \label{fig:1.9}
    \end{figure}
   
    \begin{figure}
    \begin{center}
    \begin{picture}(200,200)
     \thicklines
     \put(0,100){\vector(1,0){200}}
     \put(100,0){\vector(0,1){200}}
     \put(90,88){$0$}
     \put(205,98){$x$}
     \put(98,205){$y$}
     \put(155,200){$0<t<t_c$}

     \put(97.4,97.8){$\bullet$ }
    \end{picture}
   \end{center}
   \caption[]{Pattern with one vortex for $0<t<t_c$.
        }
   \label{fig:1.10}
    \end{figure}
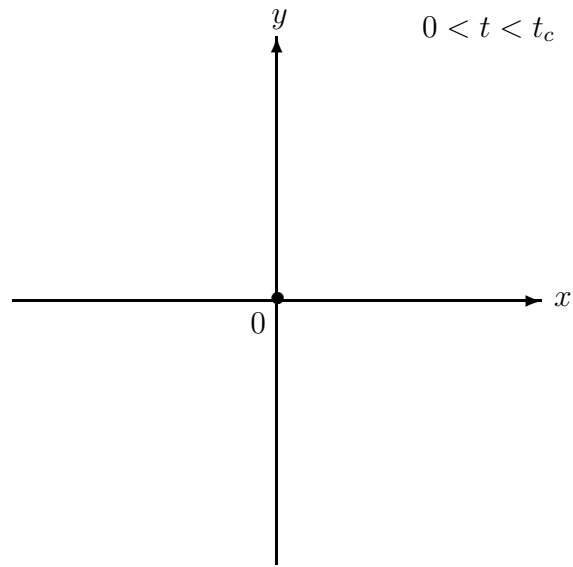

  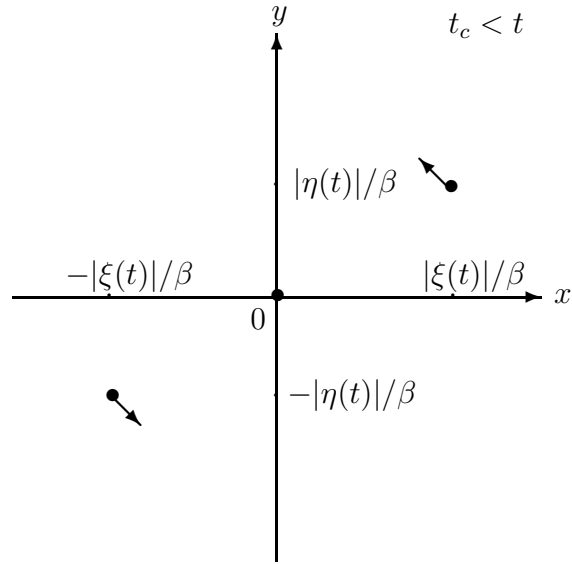
\begin{figure}
    \begin{center}
    \begin{picture}(200,200)
     \thicklines
     \put(0,100){\vector(1,0){200}}
     \put(100,0){\vector(0,1){200}}
     \put(90,88){$0$}
     \put(205,98){$x$}
     \put(98,205){$y$}
     \put(165,200){$t_c<t$}
     
     \put(165,98){$\cdot $}
     \put(35,98){$\cdot $}
     \put(155,105){$|\xi(t)|/\beta $}
     \put(20,105){$-|\xi(t)|/\beta $}
     \put(98,140){$\cdot $}
     \put(98,60){$\cdot $}
     \put(106,140){$|\eta(t)|/\beta $}
     \put(104,60){$-|\eta(t)|/\beta $}

     \put(163,139){$\bullet$}
     \put(97.4,97.8){$\bullet$ }
     \put(35,60){$\bullet $}
     \put(164,142.7){\vector(-1,1){10}}
     \put(38.5,61.5){\vector(1,-1){10}}
    \end{picture}
   \end{center}
   \caption[]{Pattern with two moving vortices for $t_c<t$ in the case of $a>0$.
       }
   \label{fig:1.11}
    \end{figure}
   
     \begin{figure}
    \begin{center}
    \begin{picture}(200,200)
     \thicklines
     \put(0,100){\vector(1,0){200}}
     \put(100,0){\vector(0,1){200}}
     \put(90,88){$0$}
     \put(205,98){$x$}
     \put(98,205){$y$}
     \put(165,200){$t\rightarrow \infty$}
     
     \put(148,98){$\cdot $}
     \put(50,98){$\cdot $}
     \put(140,105){$|c|/\beta $}
     \put(35,105){$-|c|/\beta $}
     \put(98,147){$\cdot $}
     \put(98,50){$\cdot $}
     \put(106,150){$|c|/\beta $}
     \put(104,50){$-|c|/\beta $}

     \put(147,147){$\bullet$}
     \put(97.4,97.8){$\bullet$ }
     \put(50,50){$\bullet $}
    \end{picture}
   \end{center}
   \caption[]{Recovery of the initial vortex pattern in the limit $t\rightarrow \infty$.
        }
   \label{fig:1.12}
    \end{figure}
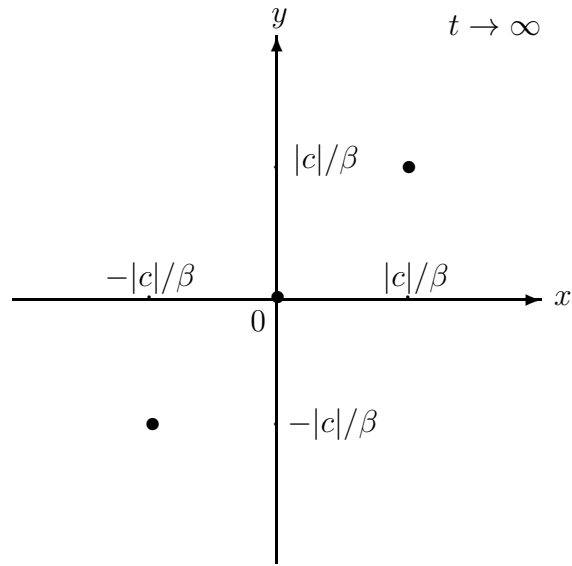

       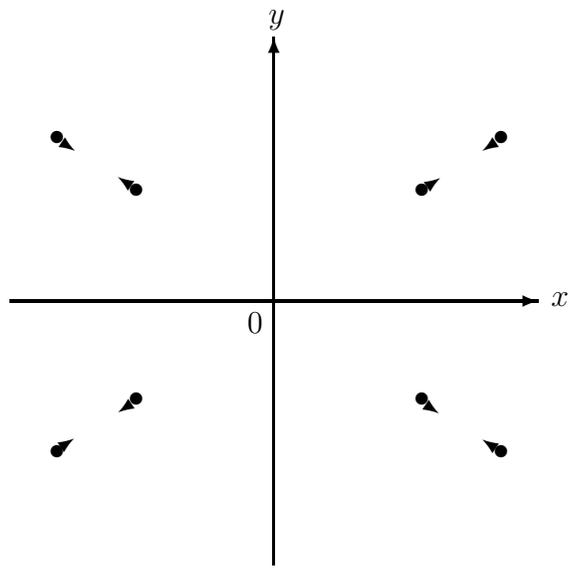
\begin{figure}
    \begin{center}
    \begin{picture}(200,200)
     \thicklines
     \put(0,100){\vector(1,0){200}}
     \put(100,0){\vector(0,1){200}}
     \put(90,88){$0$}
     \put(205,98){$x$}
     \put(98,205){$y$}

     \put(183,159){$\bullet$}
     \put(15,40){$\bullet $}
     \put(185.9,161.5){\vector(-3,-2){7}}
     \put(17.5,43.3){\vector(3,2){7}}
     \put(153,139){$\bullet$}
     \put(45,60){$\bullet $}
     \put(156,141.8){\vector(3,2){7}}
     \put(48.2,62.4){\vector(-3,-2){7}}

     \put(15,159){$\bullet$}
     \put(153,60){$\bullet $}
     \put(17.9,161.7){\vector(3,-2){7}}
     \put(155.5,62.3){\vector(3,-2){7}}
     \put(45,139){$\bullet$}
     \put(183,40){$\bullet $}
     \put(48.1,142.1){\vector(-3,2){7}}
     \put(185.9,43.0){\vector(-3,2){7}}
     
    \end{picture}
   \end{center}
   \caption[]{Four pairs creation and annihilation. 
        }
   \label{fig:1.13}
    \end{figure}


\begin{thebibliography}{99}


 \bibitem{blat}
	 G. Blatter {\it et al.},  
	 Rev. Mod. Phys. {\bf 66}, 1125 (1994). 
	 
 \bibitem{crab}
	 G. W. Crabtree and D. R. Nelson,  
	 Phys. Today {\bf 50}, No.4 38 (1997). 
 

  \bibitem{pg}
          R. E. Prange and M. Girvin M, 
          \emph{The Quantum Hall Effect} 
          (Springer, New york, 1990), 2nd ed. 
          
  \bibitem{wilczek}
          F. Wilczek, 
          \emph{Fractional Statistics and Anyon Superconductivity}
          (World Scientific, Singapore, 1990). 
          
  \bibitem{cp}
          T. Chakaraborty and P. Pietil\"{a}inen, 
          \emph{The Quantum Hall Effects: Fractional and 
          Integral} 
          (Springer, New York, 1995), 2nd and updated ed. 
          
  \bibitem{dsp}
          S. Das Sarma and A. Pinczuk,  A eds 1997 
          \emph{Perspectives in Quantum Hall Effects}
          (Wiley, New York, 1997). 
          
  \bibitem{khare}
          A. Khare, 
          \emph{Fractional Statistics and Quantum Theory}
          (World Scientific, Singapore, 1997). 
 
 \bibitem{fine}
	 K. S. Fine {\it et al.}, 
	 Phys. Rev. Lett. {\bf 75}, 3277 (1995). 

 \bibitem{kiwa1}
	 Y. Kiwamoto {\it et al.},  
	 J. Phys. Soc. Jpn. (Lett.) {\bf 68}, 3766 (1999). 

 \bibitem{kiwa2}
	 Y. Kiwamoto {\it et al.}, 
	 Phys. Rev. Lett. {\bf 85}, 3173 (2000). 
	 
 \bibitem{kiwa3}
	 K. Ito {\it et al.},  
	 Jpn. J. Appl. Phys. {\bf 40A}, 2558 (2001). 


 \bibitem{matt}
	 M. R. Matthews {\it et al.},  
	 Phys. Rev. Lett. {\bf 83}, 2498 (1999). 

 \bibitem{rama}
	 C. Raman {\it et al.}, 
	 Phys. Rev. Lett. {\bf 83}, 2502 (1999). 

 \bibitem{madi}
	 K. W. Madison {\it et al.}, 
	 Phys. Rev. Lett. {\bf 84}, 806 (2000). 

 \bibitem{mara} 
	 O. M. Marago {\it et al.}, 
	 Phys. Rev. Lett. {\bf 84}, 2056 (2000). 

 \bibitem{fitz} 
	 R. Fitzgerald {\it et al.}, 
	 Phys. Today {\bf 55}, No.8 19 (2000). 

	 
 \bibitem{lamb}
	 H. Lamb, 
	 \emph{Hydrodynamics} 
	 (Cambridge Univ. Press, Cambridge, 1932), 
	 6th ed. 
	 
 \bibitem{landau2}
	 L. D. Landau and E. M. Lifshitz, 
	 \emph{Fluid Mechanics}  
	 (Pergamon, Oxford, 1987), 
	  2nd ed.
	 
 \bibitem{batchelor}
	 G. K. Batchelor,  
	 \emph{An Introduction to Fluid Dynamics} 
	 (Cambridge Univ. Press, Cambridge, 1967). 
	 
 \bibitem{tatsu}
          T. Tatsumi, 
          \emph{Hydrodynamics} (in Japanese) 
          (Tokyo: Baihuukann 1982).
 
	 
 \bibitem{saff}
	 P. G. Saffman, 
	 \emph{Vortex Dynamics} 
	 (Cambridge Univ. Press, Cambridge, 1992). 
 \bibitem{madelung}
	 E. Madelung, 
	 Z. Phys. {\bf 40}, 322 (1926). 
  
 \bibitem{kennard}
	 E. H. Kennard,  
	 Phys. Rev. {\bf 31}, 876 (1928). 
  
 \bibitem{debroglie}
	 L. de~Broglie, 
	 \emph{Introduction \`{a} l'\'{e}tude de la 
	 M\'{e}canique ondulatoire} 
	 (Hermann, Paris, 1930). 
	 
 \bibitem{dirac51:4-54:1}
	 P. A. M. Dirac,  
	 Proc. R. Soc. London, Ser. A {\bf 209}, 291 (1951); 
	 {\it ibid.} {\bf 212}, 330 (1952); 
	 {\it ibid.} {\bf 223}, 438 (1954).
  
 \bibitem{dbohm}
	 D. Bohm,  
	 Phys. Rev. {\bf 85}, 166, 180 (1952); 
	 {\it ibid.} {\bf 89}, 458 (1953). 
  
 \bibitem{takabayasi}
	 T. Takabayasi, 
	 Prog. Theor. Phys. {\bf 8}, 143 (1952); 
	 {\it ibid.} {\bf 9}, 187 (1953). 
  
 \bibitem{schonberg}
	 M. Sch\"{o}nberg,  
	 Nuovo Cimento Soc. Ital. Fis., {\bf 12}, 103 (1954). 
  
 \bibitem{dbohm2}
	 D. Bohm and J. P. Vigier,  
	 Phys. Rev. {\bf 96}, 208 (1954); 
	 {\it ibid.} {\bf 109}, 1882 (1958). 
  
  
 \bibitem{joh1}
	 J. O. Hirschfelder, A. C. Christoph and W. E. Palke,  
	 J. Chem. Phys. {\bf 61}, 5435 (1974). 
	 
 \bibitem{joh2}
	 J. O. Hirschfelder, C. J. Goebel and L. W. Bruch,  
	 J. Chem. Phys. {\bf 61}, 5456 (1974). 
  
 \bibitem{joh3-4}
	 J. O. Hirschfelder and K. T. Tang, 
	 J. Chem. Phys. {\bf 64}, 760; 
	 {\it ibid.} {\bf 65}, 470 (1976). 
  
 \bibitem{joh5}
	 J. O. Hirschfelder,  
	 J. Chem. Phys. {\bf 67}, 5477 (1977). 
  

  \bibitem{gd}
	 S. K. Ghosh and B. M. Deb,  
	 Phys. Rep. {\bf 92}, 1 (1982). 
  
 
  
   \bibitem{wu-sp}
          H. Wu and D. W. L.Sprung, 
          \emph{Phys. Letters A} {\bf 183}, 413 (1993).
  
  \bibitem{sche}
          D. A. Schecter and H. E. Dubin, 
          \emph{Phys. Rev. Lett.} {\bf 83}, 2191 (1999). 
          
 \bibitem{bb2}
          I. Bialynicki-Birula, Z. Bialynicka-Birula 
          and \'{S}liwa~C, 
          \emph{Phys. Rev.} {\bf A61}, 032110 (2000). 
          
          
          
 
   \bibitem{k1}
          T. Kobayashi, 
          \emph{Physica} {\bf A303} (2002) 469
          .
  \bibitem{ks8}
          T. Kobayashi and T. Shimbori, 
          \emph{Phys. Rev.} {\bf A65}, 042108 (2002). 
 


 \bibitem{bohm}
	  A. Bohm and M. Gadella, 
	  \emph{Dirac Kets, Gamow Vectors and Gel'fand Triplets} 
	  (Lecture Notes in Physics, Vol. 348, Springer, 1989). 
	  
  
  \bibitem{sk4}
          T. Shimbori and T. Kobayashi, 
          \emph{J. Phys.} {\bf A33}, 7637 (2000).

 
 \bibitem{barton}
	 G. Barton, 
	 Ann. Phys. (N.Y.) {\bf 166}, 322 (1986). 
  
 \bibitem{bcd}
	 P. Briet, J. M. Combes and P. Duclos,  
	 Commun. Partial Diff. Eqns. {\bf 12}, 201 (1987).
  
 \bibitem{bv}
	 N. L. Balazs and A. Voros, 
	 Ann. Phys. (N.Y.) {\bf 199}, 123 (1990). 
  
 \bibitem{cdlp}
	 M. Castagnino, R. Diener, L. Lara and G. Puccini, 
	 Int. J. Theor. Phys. {\bf 36}, 2349 (1997). 
  
 \bibitem{sk}
	 T. Shimbori and T. Kobayashi,  
	 Nuovo Cimento Soc. Ital. Fis., B {\bf 115}, 325 (2000). 
  
 \bibitem{s2}
	 T. Shimbori, 
	 Phys. Lett. A {\bf 273}, 37 (2000). 
  
 
 \bibitem{child1} 
  M. S. Child, Proc. Roy. Soc. (London) {\bf A292} (1966) 272. 

 \bibitem{child2}
  M. S. Child, Mol. Phys. {\bf 12} (1967) 401.

 \bibitem{connor}
  J. N. L. Connor, Mol. Phys. {\bf 15} (1968) 37. 
   
  
  
  
  
  
 
  
  
  \bibitem{ks5}
          T. Kobayashi and T. Shimbori, 
          Statistical mechanics 
          for states with complex eigenvalues 
          and quasi-stable semiclassical systems 
          \emph{Preprint} cond-mat/0005237 (2000).

  \bibitem{ks6}
          T. Kobayashi and T. Shimbori, 
          \emph{Phys. Lett.} {\bf A280}, 23 (2001).
          
 \bibitem{ks7}
          T. Kobayashi and T. Shimbori, 
          \emph{Phys. Rev.} {\bf E63}, 056101 (2001). 
          
          
 
          

\end{thebibliography}
\end{document}